\documentstyle[11pt]{article}
\begin{document}
\begin{titlepage}
\vspace{0.5cm}
\centerline{\Large \bf APPLICATIONS OF THE SUPERFLAVOR SYMMETRY TO
}
\vspace{0.5cm}
\centerline{\Large \bf HEAVY BARYON-ANTIBARYON PAIR PRODUCTION}
\vspace{0.5cm}
\centerline{\Large \bf IN ELECTRON-POSITRON COLLISION
\footnote{This work was
partly supported by the National
Natural Science Foundation of China (NSFC)}}
\vspace{1cm}
\begin{center}
{\small
Xin-Heng Guo $^{a,b}$, Hong-Ying Jin $^{d}$ and Xue-Qian Li $^{c,d,e}$

\vspace{0.1cm}
a: Institute of Physics, Academia Sinica, Taipei, Taiwan, China.

\vspace{0.1cm}

b: Institute of High Energy Physics,

\vspace{0.1cm}

P.O.Box 918-4, Beijing 100039, China

\vspace{0.1cm}

c: The China Center for Advanced Science and Technology, (CCAST),

\vspace{0.1cm}

(World Laboratory)

\vspace{0.1cm}
P.O.Box 8730, Beijing, 100080, China.

\vspace{0.1cm}

d: High Energy Group, ICTP, P.O. Box 586, Trieste, 34100, Italy

\vspace{0.1cm}

e: Department of Physics,
\vspace{0.1cm}

Nankai University, Tianjin, 300071, China.}
\vspace{1cm}
\end{center}

\vspace{2cm}

\centerline{\bf Abstract}
\vspace{0.1cm}

\noindent
The baryons containing two heavy quarks and a light quark are believed to have
the diquark-quark structure and the diquark is composed of the two heavy quarks
which is a
spin-0 or spin-1 object. The superflavor symmetry
can associate productions of such heavy baryon-antibaryon pair with the heavy
meson productions. The whole scenario is presented in some details in this work,
and the observation prospect in future experiments is discussed.

\end{titlepage}

\baselineskip 18pt

\centerline{\bf\large Applications of the Superflavor Symmetry to Heavy Baryon-Antibaryon}
\vspace{0.2cm}

\centerline{\bf\large Pair Production in Electron-Positron Collisions}

\vspace{1cm}

\noindent {\bf I. Introduction}

\vspace{0.2cm}

To evaluate the hadronic exclusive processes is very difficult, because the 
hadronization is fully non-perturbative and, so far there
is not a trustworthy way to deal with it from any underlying principles. 
The meson case has been studied for many
years and remarkable progress is achieved. Since there is a good amount of
data on meson production and decay, some relatively reasonable theories
have been
developed, for example the QCD sum rules \cite{shifman}, chiral
Lagrangian \cite{geor1}
and the potential model \cite{lucha} etc.. By contrast, the baryon case is much
more complicated and obscure. It is not only because there are three valence
quarks in baryons, while only a quark-antiquark pair in mesons, but also lack 
of data and available effective theories. It is known that applications
of the QCD sum rules to the baryon case is much harder than to the meson 
case.

However, study on baryon physics can much enrich our knowledge of hadron
structure and the mechanisms in the production and decay processes.

Fortunately, the heavy quark effective theory provides a way to appropriately
simplify the evaluation of the
hadronic matrix elements \cite{isgur}\cite{geor2}. In the effective theory
extra symmetries $SU_s(2)\times SU_f(2)$ manifest and the non-perturbative effects
are attributed into the well-defined Isgur-Wise function $\xi(v\cdot v')$, where
$v$ and $v'$ are the four-velocities of the concerned heavy quarks. In the heavy
quark effective theory $P_{\mu}=m_Qv_{\mu}+p_{\mu}$ where $p$ is the
residual momentum and of the $\Lambda_{QCD}$ energy scale. Under the heavy
quark limit, $P_{\mu}=m_Qv_{\mu}$. The higher order $1/m_Q$ corrections to
some processes have also been considered \cite{mannel}.

In this work we study the processes of $e^+e^-\rightarrow X_s\overline X_{s'}$
where $X$ stands for baryons containing two heavy quarks and the subscripts
$s$ and $s'$ denote their spins. Concretely, we evaluate 
$$e^+e^-\rightarrow
X^S_{1/2}\overline X^S_{1/2},\; X^A_{1/2}\overline X^A_{1/2},\; X^A_{3/2}
\overline X^A_{3/2},\; X^A_{1/2}\overline X^A_{3/2}$$
separately, where the superscript S and A describe the diquark spin-status
(see below).

The baryon which contains two heavy quarks is believed to have the diquark
structure, namely the two heavy quarks constitute a relatively stable 
object, i.e., the diquark, while the left-over light quark moves in the color
field induced by the diquark. In this scenario, the three-body problem becomes
a simpler two-body problem, therefore a theoretical evaluation on its
properties is simplified. Falk et al., \cite{falk} investigated the heavy quark
fragmentation of baryons containing two heavy quarks in this framework. 
Since both $Q$ and $Q'$ are heavy, they are bound into diquark with a radius
much less than $1/\Lambda_{QCD}$, in the heavy quark limit, the diquark 
is a point-like color-triplet object when seen by the light quark system, 
one can expect that the 
deviation from this picture is proportional to $(\Lambda_{QCD}/m_Q)^n$ with
$n\geq 1$. Combining  a light quark with this diquark makes a baryon.

In this picture, the structure is in close analog to the heavy meson case where a heavy
quark is bound with a light quark into a color singlet. The only difference 
is the color factor. According to the authors of ref.\cite{falk}, it is an
overall normalization factor $\sqrt 3/2$.  In fact,
the source of the confinement in the potential model is not clear yet, for 
example, whether it is a scalar or vector confinement is still in dispute.
So in our work in addition to the color factor we set  a free parameter 
$\beta_1$ to the confinement $\kappa r$ term while keeping the Coulomb
piece corrected by the pure color factor, i.e., $\beta_2=0.5$,
since it is caused by the single-gluon exchange. In later numerical evaluation
we choose two typical values for $\beta_1$ (the notations will be defined
in detail in next section).

Due to the analogue of heavy meson and baryon in the diquark picture, we may 
employ the superflavor symmetry to evaluate the baryon-antibaryon production.
The superflavor symmetry was established by Georgi and Wise \cite{geor3} for
interchanging the heavy quark and scalar degrees of freedom, while Carone
\cite{carone} generalized it to the symmetry of interchanging heavy quark and
vector degrees of freedom.

Obviously, the ground state diquark can be either a color triplet scalar or
vector. They have different effective vertex form factors, so the resultant
production cross sections of the baryon-antibaryon pairs which are composed
of scalar or vector diquark are also different. In this work, we study the
cross sections of various baryon-antibaryon pair production and
compare them with
the heavy meson production rates which were estimated by several authors in the
heavy quark effective theory \cite{cohen}.

The general forms for the production amplitudes have been given by Gerogi,
Wise and Carone. However, to evaluate the concrete processes, one need to
derive the form factors for the effective vertices in $e^+e^-\rightarrow
\chi\chi^{\star}$ where $\chi$ is a scalar or vector diquark in a reliable
framework. In ref.\cite{falk}, the authors determined that in a hydrogen-like
potential, the radial wavefunction $R(0)$ is proportional to
$(C_F\alpha_s)^{3/2}$ and then the diquark production form factor 
is associated  with the zero-point wavefunction of $J/\psi$ as
$$|R_{cc}(0)|^2\approx |R_{\psi}(0)|^2/8 .$$

In our work, the situation is somewhat different, namely, we are dealing with
exclusive processes compared to ref.\cite{falk} where only inclusive processes
are
involved. Thus we employ the B-S equation to calculate the vertex form factors
under some reasonable approximations.

For the electron-positron annihilation processes, there may be two
intermediate vector-bosons, the photon and $Z^0-$boson.
In the present work we are discussing
the processes near the production thresholds of the heavy b- and c-baryons,
the concerned energy $\sqrt s$ is much lower than the $Z^0-$mass,  the
propagator ${1\over s-M_Z^2}$ is much more suppressed than ${1\over s}$,
so that the contribution of $Z^0-$boson as intermediate boson can be
neglected. This is consistent with the situation for the energy ranges
of the proposed charm-tau factory and
the B-factory. In the following we only consider the electromagnetic
interaction vertices.

The paper is organized as following. After this introduction, we
briefly discuss the diquark-quark structure of baryons, in the third section
we give a detailed derivation of the form factors of the effective vertices,
in the fourth section, in terms of the superflavor symmetry, we obtain the
baryon  production cross sections of $e^+e^-\rightarrow X\overline X$
where $X$ contains two heavy quarks. Finally the last section is
devoted to numerical evaluation and discussions.\\

\noindent{\bf II. The baryon quark structure}

\vspace{0.2cm}

In this work, we only concern the baryons which contain two heavy quarks and 
the two heavy quarks consist of a color-triplet diquark. 
The baryon quark structure has been discussed by some authors \cite{isgur1}
\cite{korn}. Meanwhile the diquark structure of baryons has also been
studied for quite a long time\cite{rich}. One may believe that there is a 
dynamical mechanism which binds two quarks into a diquark system, but
not only due to some group theory tricks.

For a all-light quark system or one heavy and two light quark system, 
because the light quarks are relativistic and the 
binding energy is not too large, whether the diquark exists as a whole
object is suspicious, at least 
the spin interactions between the soft gluons and the diaquark
do not decouple. In the previous
work, we studied the baryon which was a system of one heavy and two light quarks and
suggested that the heavy one attracts one of the light quarks to constitute
a diquark and another moves around \cite{guo}, but in that scenario
the spin interaction between the 
diquark and the light system does not decouple, so makes the application of
HQET not as reliable as the case in this work. 

By contrast, in the heavy quark effective theory, the diquark structure 
of the baryons containing two heavy quarks appears more
realistic and convincing. The two heavy
quarks constitute a stable diquark, at least at the heavy quark
limit. In this scenario, since $b$ and $c$ are isospin singlets, so by
group theory, one can construct seven different baryons which contain a
heavy diquark and a light quark.
\begin{eqnarray}
X_{[b,c]_1}(1/2,1/2) &=& {1\over 3}\epsilon_{ijk}[b^{\dagger}_i\uparrow
c^{\dagger}_j\uparrow q^{\dagger}_k\downarrow-{1\over 2} b^{\dagger}_i\uparrow
c^{\dagger}_j\downarrow q^{\dagger}_k\uparrow-{1\over 2} b^{\dagger}_i\downarrow
c^{\dagger}_j\uparrow q^{\dagger}_k\uparrow]|0> \\
X_{[b,c]_0}(1/2,1/2)&=&{1\over\sqrt{12}}\epsilon_{ijk}[b^{\dagger}_i\uparrow
c^{\dagger}_j\downarrow q^{\dagger}_k\uparrow -b^{\dagger}_i\downarrow
c^{\dagger}_j\uparrow q^{\dagger}_k\uparrow]|0> \\
X_{cc}(1/2,1/2) &=& {1\over 3}\epsilon_{ijk}[c^{\dagger}_i\uparrow
c^{\dagger}_j\uparrow q^{\dagger}_k\downarrow-{1\over 2} c^{\dagger}_i\uparrow
c^{\dagger}_j\downarrow q^{\dagger}_k\uparrow-{1\over 2} c^{\dagger}_i\downarrow
c^{\dagger}_j\uparrow q^{\dagger}_k\uparrow]|0> \\
X_{bb}(1/2,1/2)&=&{1\over 3}\epsilon_{ijk}[b^{\dagger}_i\uparrow
b^{\dagger}_j\uparrow q^{\dagger}_k\downarrow-{1\over 2} b^{\dagger}_i\uparrow
b^{\dagger}_j\downarrow q^{\dagger}_k\uparrow-{1\over 2} b^{\dagger}_i\downarrow
b^{\dagger}_j\uparrow q^{\dagger}_k\uparrow]|0> \\
X_{cc}(3/2,3/2)&=&{1\over\sqrt 6}\epsilon_{ijk}[c^{\dagger}_i\uparrow c^{\dagger}_j
\uparrow q^{\dagger}_k\uparrow]|0> \\
X_{bb}(3/2,3/2)&=&{1\over\sqrt 6}\epsilon_{ijk}[b^{\dagger}_i\uparrow b^{\dagger}_j
\uparrow q^{\dagger}_k\uparrow]|0> \\
X_{bc}(3/2,3/2)&=&{1\over\sqrt 6}\epsilon_{ijk}[b^{\dagger}_i\uparrow c^{\dagger}_j
\uparrow q^{\dagger}_k\uparrow]|0>
\end{eqnarray}
where the $q^{\dagger}$'s are creation operators of quarks, i, j, k are the color
indices, the subscripts $cc$, $bc$ and $bb$ denote the heavy quark constituents in
the baryon, due to the Pauli principle, $cc$ and $bb$ can only be in the spin-1
state while $bc$ can be in either spin-1 or spin-0 states, namely diquarks $bb$ and
$cc$ are color triplet spin-1 diquarks while $bc$ can be color triplet spin-1
$\chi_{[bc]_1}$ or scalar $\chi_{[bc]_0}$. In this work we only concern 
the ground state
baryons, so the baryons can only be spin-1/2 and 3/2. The excited baryon states
were discussed by K\"{o}rner et al. \cite{korn}.

As a matter of fact, when the two quarks are heavy, we have the superflavor
symmetry which relates the low energy matrix elements of heavy mesons
and baryons with two heavy quarks. The production of heavy diquarks
from a virtual photon has effective form factors $g_A$ or $g_S$ (see
section III) which is factorized out of the low energy matrix elements.
We derive these form factors for the effective production
vertex from the B-S equation.\\

\noindent{\bf III. Derivation of the effective vertices for production of
color triplet spin-1 and -0 diquarks}

\vspace{0.2cm}

The physical picture for the production of color triplet spin-1 or -0 diquarks
is similar to that of $Z^0$ decay into charmonium via charm quark fragmentation
described by several authors \cite{brat} and the inclusive diquark production
was also estimated \cite{han}. Later Falk et al., employed this picture to
describe inclusive baryon production with the diquark picture. Here in this
work we are going to deal with the exclusive processes of baryon production
in $e^+e^-$ collisions with a similar approach.

It is noted that for applying the HQET the diquark should be of a point-like
structure, the reason is that all nonperturbative effects are attributed
into a well-defined Isgur-Wise function, therefore the necessary condition
is that the diquark is seen by the light quark as point-like. However, it
by no means demands that electromagnetic current would see a point-like
structureless object, by contraries, there is a complicated structure, but
derivable in the framework of perturbative QCD theory, so the virtual photon
would "see" an effective vertex and deriving it is the task of this section.

The Feynman diagrams are shown in Fig.1 where the off-shell photon produces
a pair of heavy quark-antiquark, then an off-shell gluon is emitted from one
leg and turns into another pair of heavy quark-antiquark. The produced
quarks and antiquarks are bound into a diquark and an antidiquark. Finally,
they pick up a light quark and a light  antiquark respectively to constitute
a baryon-antibaryon pair. Since it is very difficult to pick up a heavy
quark from the sea, the contribution from the current which couples to the light
quark-antiquark can be ignored.
It is also noticed that if $Q$ and $Q'$ are different
heavy quarks ($b$ and $c$ explicitly), there are four topologically
distinct Feynman diagrams corresponding to (a) through (d) in Fig.1, while
$Q$ and $Q'$ are the same quarks ($bb$ or $cc$), there only two
diagrams (a) and (b) exist.

Since the emitted gluon turns to heavy quark-antiquark, the energy scale for
this process is large, so the gluon is hard. At this energy scale, the
perturbative QCD reliably applies. Therefore the form factors derived
in the framework of perturbative QCD make sense. 
Hence in our case,
one can employ the perturbative QCD confidently and only needs to consider 
the leading order Feynman diagrams which are shown in Fig.1.

In this section in terms of the B-S equation \cite{bethe}, we derive the
form factors of the effective vertices for the diquark-antidiquark production
and then we evaluate their numerical results.

The B-S equation of a diquark can be written in the following form
\begin{equation}
\label{ji1}
\chi_P(p)=S_1(\lambda_1P+p)\int G(P,p,q)\chi_P(q)
{d^4q\over (2\pi)^4}S_2(\lambda_2P-p),
\end{equation}
where $S_i(i=1,2)$ are the propagators of quark 1 and quark 2 in the diquark
respectively and $G(P,p,q)$ is the reductive kernel, $\lambda_1={m_1\over
m_1+m_2}$, $\lambda_2={m_2\over m_1+m_2}$, and $m_1$, $m_2$ are the quark 
masses. $P$ is the total momentum of the diquark and can be expressed as 
$P=Mv$ where $M$ is the mass of the diquark and $v$ is its four-velocity.

Using the relation
\begin{equation}
\label{ji2}
S_j(p)=i[{\Lambda^+_j(p_t)\over p_l-W_j+i\epsilon}+
{\Lambda^-_j(p_t)\over p_l+W_j-
i\epsilon}]\rlap/v \;\;\;\;\; (j=1,2)
\end{equation}
where $p_l=p\cdot v$, $p_t=p-p_lv$, $W_j=\sqrt {|p_t|^2+m_j^2}$ and 
$\Lambda^{\pm}_j(p_t)={W_j\pm\rlap/v (-p_t+m_j)\over 2W_j}$, eq.(\ref{ji1})
can be expressed explicitly as
\begin{eqnarray}
\chi^{++}_P(p) && = {-\Lambda_1^+(p_t)\rlap/v\over \lambda_1M+p_l-W_1
+i\epsilon}
\int G(P,p,q)[\chi^{++}(q)+\chi^{--}(q)]{d^4q\over (2\pi)^4}  \nonumber \\
&& {\rlap/v\Lambda^+_2(-p_t)\over p_l+W_2-\lambda_2M-i\epsilon} \\
\chi^{--}_P(p) && = {-\Lambda_1^-(p_t)\rlap/v\over \lambda_1M+p_l+W_1-i\epsilon}
\int G(P,p,q)[\chi^{++}(q)+\chi^{--}(q)]{d^4q\over (2\pi)^4} \nonumber \\
&& {\rlap/v\Lambda^-_2(-p_t)\over p_l-W_2-\lambda_2M+i\epsilon}
\end{eqnarray}
where $\chi^{\pm\pm}_P(p)=\Lambda^{\pm}_1(p_t)\chi_P(p)\Lambda^{\pm}_2(-p_t)$.

In the non-relativistic approximation, which  applies to the low-lying states
of the two-heavy-quark system, $\chi_P^{--}$ is small and negligible at
the first order
and $\Lambda_1^+(p_t)\approx {1+\rlap/v \over 2}$, $\Lambda_2^+(-p_t)
\approx {1+\rlap/v\over 2}$.

So for a  scalar or an axial vector diquark, the B-S wavefunction can be written
in the forms
$$\chi^S_P(p)={1+\rlap/v\over 2}\sqrt{2M}\phi(p),\;\;\;{\rm or}\;\;\;
\chi^A_P(p)={1+\rlap/v\over 2}\sqrt{2M}\gamma_5\rlap/{\eta}\phi(p).$$
The superscript S and A denote the scalar and axial vector 
respectively and $\eta$ is the polarization vector of the vector-diquark.

Now we assume the kernel $G$ to have a form
\begin{equation}
-iG=1\otimes 1 V_1+\rlap/v\otimes\rlap/v V_2 
\end{equation}
where 
$$V_1(p,q)={8\pi \beta_1\kappa\over [(p_t-q_t)^2+\mu^2]^2}-
(2\pi)^3\delta^3(p_t-q_t) \displaystyle\int{\frac{8\pi\beta_1\kappa}
{(k^2+\mu^2)^2}\frac{d^3k}{(2\pi)^3}} $$
and 
$$V_2(p,q)=-\displaystyle{\frac{16\pi\beta_2\alpha_s}{3(|p_t-q_t|^2+\mu^2)}}$$
The parameters $\beta_1$ and $\beta_2$ are different for the  
various color states. For mesons, $\beta_1=1$, $\beta_2=1$, while for 
color-triplet diquarks,  
$\beta_2$ is directly associated to the color factor caused by the single-gluon
exchange, so should be 0.5, in contrast, $\beta_1$ which is related to
the linear confinement, as aforementioned, cannot be determined so far and 
we just take it as a free parameter in numerical evaluations. 
As a matter of fact, later
we pick up two typical values 0.5 and 1 for $\beta_1$ for demonstrating the
influence of the color factor. The parameters $\kappa$ and $\alpha_s$ are
well determined by fitting  experimental data of heavy meson spectra. 
From the heavy meson
experimental data,  $\kappa=0.18$, $\alpha_s=0.4$.\cite{eich}
Then, we solve the integrational equation
\begin{equation}
{\tilde \phi}(p_t)=\displaystyle{\frac{-1}{M-W_1-W_2}\int(V_1-V_2)
{\tilde \phi}(q_t)\frac{d^3q_t}{(2\pi)^3}} 
\end{equation}
to obtain the B-S wave function by numerical calculation, where  
${\tilde\phi}(p_t)=\int \phi(p)\frac{dp_l}{2\pi}$.

In general, the matrix elements of the diquark-antidiquark production  
from electromagnetic interaction can be factorized as  
\begin{eqnarray}
\label{ji5}
&& <M^S(v')M^{* S}(v)|J^{\mu}|0>=Mf_1(v\cdot v')(v'-v)^\mu, \nonumber\\
&&<M^A(v',\eta ')M^{* A}(v,\eta )|J^\mu|0>=
M[f_2(v\cdot v')\eta '\cdot\eta (v'_\mu-v_\mu) 
+ f_3(v\cdot v')(\eta\cdot v' \eta_\mu'-\eta '\cdot v\eta_\mu)]. \nonumber \\
\end{eqnarray}

On the other hand, the effective matrix elements of Fig.1 can be expressed
by the B-S wave function in following 
\begin{equation}
<MM^{\star}|J|0>=\int
Tr[{\bar \chi}^{M}_{P'}(p')(\Omega_1+\Omega_2+\Omega_3+\Omega_4)
\chi^{M^\star}_P(p)\displaystyle{\frac{d^4p}{(2\pi)^4}\frac{d^4p'}{(2\pi)^4}}]
\end{equation}
where $J=\sum_{k=1}^2\bar h_k\Gamma_kh_k$, $M^{\star}$ is the anti-particle of M and $\chi_P^{\star S}=
\frac{1-\rlap/v}{2}\sqrt{2M}\phi(p)$, $\chi_P^{\star A}=\frac{1-\rlap/v} 
{2}\gamma^5\rlap/\eta\phi(p)$ for scalar- and axial vector-state respectively.
The $\Omega_i$(i=1,..,4) are defined as
\begin{equation}
\begin{array}{l}

\Omega_1=-{2\over 3}g^2_s\Gamma_1 S_1(p'_1-q)\gamma^\mu\otimes\gamma_\mu D(p'_1-p_1-q),\\
\Omega_2=-{2\over 3}g^2_s\gamma^\mu S_1(p_1+q)\Gamma_1\otimes\gamma_\mu D(p'_1-p_1-q),\\  
\Omega_3=-{2\over 3}g^2_s\gamma^\mu\otimes\gamma_\mu S_2(p'_2-q)\Gamma_2 D(p'_2-p_2-q),\\ 
\Omega_4=-{2\over 3}g^2_s\gamma^\mu\otimes\Gamma_2 S_2(p_2+q)\gamma_\mu D(p'_2-p_2-q)\\
\end{array}
\end{equation}
respectively, where $g^{\mu\nu}D$ is the propagator of the gluon, $g_s$ is    
the coupling constant of strong interaction, $p'_i$ and  
$p_i$(i=1,2)  are  
\begin{equation}
\begin{array}{ll}
p'_1=\lambda_1 Mv'+p',~~~~~~~~&p'_2=\lambda_2 Mv'-p'\\
p_1=-\lambda_1 Mv-p ,~~~&p_2=-\lambda_2 Mv+p\\
\end{array}
\end{equation}
respectively. 

Similar to ref.\cite{gube}, we ignore the dependence of $p'$ and $p$ in $\Omega_i$
(i=1,..,4), then for electromagnetic current we find
\begin{eqnarray}
\label{fex}
&& f_1=(Q_1E(m_1,m_2)+Q_2E(m_2,m_1)), \nonumber\\
&& f_2=\displaystyle{(-Q_1f(m_1,m_2)-Q_2f(m_2,m_1))}, \nonumber\\
&& f_3=\displaystyle{(Q_1g(m_1,m_2)+Q_2g(m_2,m_1))}, 
\end{eqnarray}
where $Q_ie$ is the charge of quark i (i=1,2) respectively and
\begin{eqnarray}
\label{fex1}
E(m_1,m_2)&=&\displaystyle{[\frac{m_1}{M}+\lambda_1-2\lambda_2(1+v\cdot v')]
h(m_1,m_2)}
\nonumber \\
&\approx&2[\lambda_1-\lambda_2(1+v\cdot v')]h(m_1,m_2)\nonumber\\
E(m_2,m_1)&=&\displaystyle{[\frac{m_2}{M}+\lambda_2-2\lambda_1(1+v\cdot v')]
h(m_2,m_1)}
\nonumber\\
&\approx& 2[\lambda_2-\lambda_1(1+v\cdot v')]h(m_2,m_1)\nonumber\\
f(m_1,m_2)&=&\displaystyle{(\frac{m_1}{M}+\lambda_1)h(m_1,m_2)} \nonumber\\
&\approx&2\lambda_1h(m_1,m_2), \nonumber\\
f(m_2,m_1)&=&\displaystyle{(\frac{m_2}{M}+\lambda_2)h(m_2,m_1)} \nonumber\\
&\approx&2\lambda_2h(m_2,m_1) \nonumber\\
g(m_1,m_2)&=&\displaystyle{(\frac{m_1}{M}+\lambda_2+1)h(m_1,m_2)} \nonumber\\
&\approx&2h(m_1,m_2), \nonumber\\
g(m_2,m_1)&=&\displaystyle{(\frac{m_2}{M}+\lambda_1+1)h(m_2,m_1)} \nonumber\\
&\approx&2h(m_2,m_1), \nonumber\\
h(m_1,m_2)&=&\displaystyle{\frac{1}{M^3\lambda_2^2(1+v\cdot v')(1+\lambda_2^2
-\lambda_1^2+2\lambda_2v\cdot v')}{2g^2_s(4m^2_2)F^2\over 3}}  \nonumber\\
&\approx&\displaystyle{\frac{1}{3(1+v\cdot v')^2}
\frac{g_s^2(4m^2_2)}{m_2^3}F^2},
\nonumber\\
h(m_2,m_1)&=&\displaystyle{\frac{1}{M^3\lambda_1^2(1+v\cdot v')(1+\lambda_1^2
-\lambda_2^2+2\lambda_1v\cdot v')}{2g^2_s(4m^2_1)F^2\over 3}} \nonumber\\
&\approx&\displaystyle{\frac{1}{3(1+v\cdot v')^2}
\frac{g_s^2(4m^2_1)}{m_1^3}F^2},
\nonumber\\
F &=& \displaystyle{\int \phi(p)\frac{d^4p}{(2\pi)^4}}.
\end{eqnarray}
For obtaining the final results of the above  expressions, the approximation
$M=m_1+m_2$ is taken.

The numerical results are shown in table 1. The masses of b-quark and c-quark
are chosen to be $m_b=5.02$ GeV, $m_c=1.58$ GeV which are determined
in ref.\cite{jin}, $\beta_1$ is to be 0.5 or
1.0 for a comparison. 

\begin{center}

\begin{tabular}{|r|r|r|r|r|r|r|}
\hline
$\beta_1$ &0.5 &1 &0.5& 1 &0.5& 1 \\
\hline
$m_1(GeV)$ & 5.02 & 5.02 & 5.02 & 5.02 &1.58 &1.58\\
\hline
$Q_1$ & -1/3 & -1/3 & -1/3 & -1/3 & 2/3& 2/3\\
\hline
$m_2(GeV)$ & 5.02 & 5.02 & 1.58 & 1.58 & 1.58 & 1.58 \\
\hline
$Q_2$ & -1/3 & -1/3 & 2/3 & 2/3 & 2/3 & 2/3\\
\hline 
$M(GeV)$ & 10.08& 10.17 & 6.76& 6.89& 3.40 & 3.57\\
\hline
$F(GeV^{3/2})$ & 0.287 &0.341 & 0.177& 0.226 & 0.128 & 0.169\\
\hline
\end{tabular}

\end{center}

\vspace{0.2cm}

\centerline {Table 1}

The concerned numerical values  where $\alpha_s=0.4$,
$\kappa=0.18$ GeV$^2$,$m_b=5.02$ GeV, $m_c=1.58$ GeV. \\

The forms of the flavor currents to produce the diquark pairs we will use
are the following 

\begin{equation}
\label{guo1}
J^{\lambda}_S=ig_S(\chi^+\partial^{\lambda}\chi-\partial^{\lambda}\chi^+\chi),
\end{equation}
\begin{equation}
\label{guo2}
J^{\lambda}_A=-ig_A[A^{\mu +}\partial_{\mu}A^{\lambda}-(\partial_{\mu}
A^{\lambda})^{+}A^{\mu}-a(A^{\mu +}\partial^{\lambda}A_{\mu}-(\partial^{\lambda}
A_{\mu})^{+}A^{\mu})], 
\end{equation}
where $J^{\lambda}_S$ is the current for scalar and $J^{\lambda}_A$ is for
axial vector
$g_S$ and $g_A$ are the
effective vertex form factors and $a$ is a parameter which can be found to be the 
minus ration of $f_2$ over $f_3$ in eq.(\ref{ji5}).

Comparing eqs.(\ref{fex}) (\ref{fex1}) (\ref{guo1}) and (\ref{guo2}) one obtains
\begin{eqnarray}
g^{bb}_A&=&\displaystyle{\frac{4e}{9(1+v\cdot v')^2}
\frac{4\pi\alpha_s(4m^2_b)}{m_b^3}F_{bb}^2}\\
g^{bc}_S&=&\displaystyle{\frac{2e}{9(1+v\cdot v')^2}[\frac{m_b}{m_b+m_c}
\frac{4\pi\alpha_s(4m^2_c)}{m_c^3}-\frac{2m_c}{m_b+m_c}\frac{4\pi
\alpha_s(4m^2_b)}
{m_b^3}]F_{bc}^2}\nonumber\\
&&-\displaystyle{\frac{2e}{9(1+v\cdot v')}
[\frac{4\pi\alpha_s(4m^2_c)}
{(m_b+m_c)m_c^2}-\frac{8\pi\alpha_s(4m^2_b)}{(m_b+m_c)m_b^2}]F_{bc}^2}\\
g^{bc}_A&=&\displaystyle{\frac{2e}{9(1+v\cdot v')^2}[\frac{m_b}{m_b+m_c}
\frac{4\pi\alpha_s(4m^2_c)}{m_c^3}-\frac{2m_c}{m_b+m_c}
\frac{4\pi\alpha_s(4m^2_b)}
{m_b^3}]F_{bc}^2}\\
g^{cc}_A&=&\displaystyle{\frac{-8e}{9(1+v\cdot v')^2}
\frac{4\pi\alpha_s(4m^2_c)}{m_c^3}F_{cc}^2},
\end{eqnarray}
where $\alpha_s(M^2)$ is the running coupling constant of QCD. 
It can be seen from eqs.(\ref{fex}) (\ref{fex1}) (\ref{guo1}) 
and (\ref{guo2}) that a=0.5
for bb and cc cases while a=1 for bc diquark.

\vspace{0.2cm}

\noindent{\bf IV. The superflavor symmetry and the production cross sections}

\vspace{0.2cm}

The superflavor symmetry interchanges the heavy quark and spin-0 or spin-1
degrees of freedom, so can determine relations between their 
hadronic matrix elements \cite{geor3}\cite{carone}.
For the scalar case the fields of the heavy quark and the scalar can be put
together into a five-component column vector with a given velocity,
\cite{geor3}
$$\Psi_v=\left( \begin{array}{c}
h^+_v \\ \chi_v \end{array} \right) $$
where $h_v^+$ is the heavy quark spinor, $\rlap/v h^+_v=h^+_v$
and $\chi_v$ is the heavy scalar field, while for the spin-1 case, the wavefunction
becomes an 8-component column vector \cite{carone},
$$\Psi_v=\left( \begin{array}{c}
h^+_v \\  A^{\mu}_v  \end{array} \right) $$
where $A^{\mu}_v$ is a heavy axial-vector field with a constraint $v_{\mu}A_v^{\mu}=0$.
One can write the wavefunctions of meson and baryon corresponding to the heavy
quark and scalar $\chi_v$ or vector $A_v^{\mu}$ diquarks as
$$ \Psi_H(v)=\left( \begin{array}{c}
\sqrt {m_h}\gamma_5{1\over 2}(1-\rlap/v) \\ 0 \end{array} \right) \;\;\;\;
\Psi_{H^*}(v)=\left( \begin{array}{c} \sqrt{m_h}\rlap/{\epsilon}{1\over 2}
(1-\gamma_5) \\ 0 \end{array} \right) $$
and
$$\Psi_{\chi_S}^{1/2}(v)=\left( \begin{array}{c} 0 \\
U^TC/\sqrt{2m_{\chi_S}} \end{array} \right) \;\;\;\; \Psi_{\chi_A}^{1/2}(v)=
{1\over \sqrt{6m_A}} \left( \begin{array}{c} 0 \\ U^TC\sigma^{\mu\beta}v_{\beta}
\gamma_5 \end{array} \right) $$
$$ \Psi_{\chi_A}^{3/2}(v)={1\over
\sqrt{2m_A}}\left( \begin{array}{c} 0 \\ U^{\mu T}C \end{array} \right) $$
where $U$ is the spinor of baryons and $C$ is the charge conjugation operator,
$U^{\mu}$ is the Rarita-Schwinger spinor-vector wavefunction satisfying
constraints $v_{\mu}U^{\mu}=0$ and $\gamma_{\mu}U^{\mu}=0$.

Below we will give the concrete forms for the production amplitudes for
$$e^+e^-\rightarrow
H\overline H, H^*\overline H^*, X_S\overline X_S, X_V(1/2)\overline X_V(1/2),
X_V(3/2)\overline X_V(3/2) \;{\rm and\;} X_V(1/2)\overline X_V(3/2), $$
where $H, H^*, X_S, X_A(1/2)$ and $X_A(3/2)$ denote the meson, vector-meson,
spin-1/2 baryon with the diquark being a scalar, spin-1/2 baryon with the
spin-1 diquark  and spin-3/2 baryon respectively.

The meson production rate was calculated \cite{cohen} as
\begin{equation}
<H(v')\overline H(v)|J_{\lambda}|0>=f_+(-v\cdot v')(P_H-
P_{\overline H})_{\lambda}.
\end{equation}

In the approach given by Georgi and Wise, \cite{geor3}
\begin{eqnarray}
&& <H(v')\overline H(v)|\bar h\gamma_{\lambda}h|0>=\xi(-v\cdot v')
m_h(v'-v)_{\lambda}, \\
&& <H^*(v')\overline H^*(v)|\bar h\gamma_{\lambda}h|0>=-\xi(-v\cdot v')
[(\epsilon'^*\cdot\epsilon)(v'-v)_{\lambda}+(\epsilon'^*\cdot v)\epsilon_{\lambda}
-(\epsilon\cdot v')\epsilon'^*],
\end{eqnarray}
where $\xi(v\cdot v')$ is the Isgur-Wise function with the normalization $\xi(1)=1$.

Similarly for the baryon case, 
\begin{eqnarray}
&& <X_S(v')\overline X_S(v)|J^{\lambda}_S|0>
=g_S\xi(-v\cdot v'){1\over 2}(v'-v)^{\lambda}\overline U'V \\
&& <X_A^{1/2}\overline X_A^{1/2}|J^{\lambda}_A|0>=
{1\over 6}g_A\xi(-v\cdot v')[a(2-v\cdot v')(v'-v)^{\lambda}\overline U'V \nonumber \\
&&+(1-v'v)\overline U'(2\gamma^{\lambda}+v^{\lambda}-v'^{\lambda})V] \\
&& <X_A^{3/2}\overline X_A^{3/2}|J^\lambda_A|0>={1\over 2}g_A
\xi(-v\cdot v')[-a(v'-v)^{\lambda}\overline U'_{\mu}V^{\mu}
+\overline U'^{\lambda}v'_{\alpha}V^{\alpha}-v_{\alpha}\overline U'^{\alpha}
V^{\lambda}], \\
&& <X_A^{1/2}\overline X_A^{3/2}|J^\lambda_A|0>={1\over 2\sqrt 3}g_A
\xi(-v\cdot v') [(1-v\cdot v')\overline U'\gamma_5 V^{\lambda}
+\overline U'\gamma_5(\gamma^{\lambda}-av^{\lambda}+(1-a)v'^{\lambda})
v'_{\alpha}V^{\alpha}, \nonumber\\
\end{eqnarray}
where $U$ and $V$ are spinors of the baryon and antibaryon, the value of 
$a$ has been discussed above.

Thus one can obtain the cross section for the productions as
\begin{equation}
\sigma={m_e^2\over {s\over 2}}{1\over 4}\int {d^3p_1\over (2\pi)^3}{M\over E_1}
{d^3p_2\over (2\pi)^3}{M\over E_2}(2\pi)^4\delta^4(p+p'-p_1-p_2)|T|^2.
\end{equation}
It is noticed that here for the cross section evaluation the normalization is
$\overline UU=1$.

Without losing generality, we set $p=({\sqrt s\over 2},
{\sqrt s\over 2}\hat z)$ as the momentum of the electron, and 
$p'=({\sqrt s\over 2}, {-\sqrt s\over 2}\hat z)$ as the momentum of positron
$e^+$.
$p_1=({\sqrt s\over 2}, {\sqrt{s-4M^2}\over 2}\hat n)$ is the 
momentum of the outgoing baryon, and in the heavy quark limit $Mv'=p_1$, while
$p_2=({\sqrt s\over 2}, {-\sqrt{s-4M^2}\over 2}\hat n)$ is the momentum
of the outgoing antibaryon, and in the heavy quark limit $Mv=p_2$.
$\hat n$ is an arbitrary  three-dimensional unit vector as $\hat n^2=1$ and
$\hat n\cdot \hat z=cos\theta$.
In the formulae, we denote the mass of the heavy baryon as M and that of 
the electron as $m_e$.

\begin{eqnarray}
\label{baryon}
&& \sigma(e^+e^-\rightarrow X_S\overline X_S)=
{e^2g_S^2\over 48\pi s^3\sqrt sM^2}\sqrt{(s-4M^2)}
|\xi(-\omega)|^2s({s\over 2}-2M^2)(s-4M^2), \\
&& \sigma(e^+e^-\rightarrow X_A^{1/2}\overline X^{1/2}_A)={e^2g_A^2\over
3456\pi s^2\sqrt sM^6}\sqrt{(s-4M^2)}|\xi(-\omega)|^2
[64(2a+1)^2M^8  \nonumber \\
&& +32(-8a^2-2a+5)M^6 s+24(4a^2-2a-3)M^4 s^2+4(-4a^2+5a+1)M^2 s^3 \nonumber \\
&& +(a-1)^2s^4], \\
&& \sigma(e^+e^-\rightarrow X_A^{3/2}\overline X_A^{3/2})={e^2g_A^2\over
864\pi s^2\sqrt sM^6}\sqrt{(s-4M^2)}|\xi(-\omega)|^2
[288a^2 M^8 \nonumber \\
&& +16(-11a^2+3a+4)M^6 s+2(25a^2-26a-6)M^4 s^2 
-(10a^2-18a+5)M^2 s^3 \nonumber \\
&& -(a-1)^2s^4],\\
&& \sigma(e^+e^-\rightarrow X_A^{1/2}\overline X_A^{3/2})={e^2g_A^2\over
1728\pi s\sqrt sM^6}\sqrt{(s-4M^2)}|\xi(-\omega)|^2
[32M^6 +16(a^2-2a)M^4 s \nonumber \\
&&+2(4a^2+8a-3)M^2 s^2 +(a-1)^2 s^3],
\end{eqnarray}
where $\omega=v\cdot v'$, $g_S$ and $g_A$ are the effective vertex form factors and
derived in last section and again a=1 for bc diquark and a=0.5 for bb or cc
diquark. Moreover, the differential cross sections
$d\sigma /dcos\theta$ are given in the appendix. It is noted that the 
differential cross sections can be grouped into the transverse piece multiplying
$(1+cos^2 \theta)$ and the longitudinal one multiplying $sin^2 \theta$ 
respectively \cite{korn1}.

In comparison with the meson production case $e^+e^-\rightarrow D\overline D$,
whose  cross section is
\begin{equation}
\label{meson}
\sigma(e^+e^-\rightarrow D\overline D)={e^2g'^2\over 6\pi s^2\sqrt s}
|\xi(-\omega)|^2({s\over 4}-M_D^2)^{3/2}
\end{equation}
where $g'={2\over 3}e$ for the charm-meson case, one can immediately obtain the
ratios.

In next section, we will discuss numerical results.\\

\noindent{\bf V. Numerical results and discussions}
\vspace{0.2cm}

Since so far, the baryons which contain two heavy quarks have not been detected
in experiments yet, one cannot determine their masses precisely. However, in
the heavy quark theory, their masses are very close to the sum of the two
heavy quarks, because the binding energy is of the $\Lambda_{QCD}$ scale which
is much smaller than the heavy quark mass. Moreover, in the heavy quark effective
theory there is an extra symmetry $SU_s(2)\times SU_f(2)$, the spin splitting
is at $1/M$ order. Therefore, in the numerical evaluation, we simply use
$$M\approx m_{Q_1}+m_{Q_2}+\Lambda$$
while the binding energy $\Lambda$ is calculated in terms of the B-S equation.

The values calculated in terms of the B-S equation are
$M_{bb}=10.17$ GeV, $M_{bc}=6.89$ GeV and $M_{cc}=3.57$ GeV.
In Tables 2,3 and 4 shown in Appendix B we give the numerical values of the
cross sections corresponding to $X_{cc}, X_{bc}$ and $X_{bb}$ respectively
in a range of $\sqrt{s}$ above the threshold which corresponds to $\omega$
from 1 to 2. In Table 5 we show the results for $\sigma (e^+e^- \rightarrow
D\overline D)$ for comparision.

All non-perturbative QCD effects associated
with light quarks are attributed to the Isgur-Wise function. 
It is noted that the cross sections are proportional to the Isgur-Wise 
function which is at a negative
argument region, $\xi(-\omega)$. Because a transition
from Q to Q' is at the s-channel which is the time-like region,
so the argument is positive $v\cdot v'$, in contrast,
the production of pair $Q\overline Q'$ is at the t-channel which is the
space-like region and the argument is $-v\cdot v'$.
The Isgur-Wise function is normalized to
$\xi(1)=1$ and can be expanded at small $\omega=v\cdot v'$\cite{fun},
but since the negative $\omega-$values are far apart from 1, so an
extrapolation is not legitimate, and to our knowledge it has not been evaluated
at the space-like regions. It is understood that the experiments to obtain 
the cross section of meson-anti-meson (for instance 
$e^+e^- \rightarrow D\overline D)$ should be much easier than those
for $X_{cc}, X_{bc}$ and $X_{bb}$ and actually BEPC is just working at this
energy region. If the cross section for meson anti-meson is measured,
with help of the superflavor symmetry scenario,  the
results can be associated with that for heavy baryon production, so we may
elude the troublesome point since the $\xi(-\omega)$ can be cancelled for
baryon and meson cases if they have the same $\omega=s/2M^2 -1$. Therefore,
if we know the cross section for, say $D \overline D$ at some $\omega$
we can give the predictions for $X_{cc}, X_{bc}$ and $X_{bb}$ at the same
$\omega$. The easiest regions are near the production thresholds where $\omega$
is near 1. In this case $\xi (-\omega)$ is not far away from the value
at $\omega =1$, hence $\xi (-\omega)$ can be cancelled in baryon and meson
cases.

In fact, just above the threshold of $D \overline D$ production, there is 
a rich spectrum structure with many resonances crowded in a small region, but
not for the diquark-antidiquark case in the present work. This is because
we are dealing with the baryon production in the leading order of HQET where
$m_Q \rightarrow \infty$ we can neglect the resonance structure in our 
calculations. The reason is that the separation of resonances are related
to $1/m_Q$ corrections which are ignored at present in our calculations.

It is noted that the heavy baryon production rates in the tables shown
in Appendix B are 5
to 8 orders smaller than the heavy meson production rates and it seems
reasonable.

The BEPC is going to be upgraded to higher energies and luminosity and the
proposed charm-tau factory is under discussion, meanwhile the B-factory
is also under way and their energy is enough to produce pairs of baryon-antibaryon
containing two heavy quarks and the luminosity of $10^{33}\; cm^{-2}s^{-1}$
as proposed can also
produce sufficient events. Therefore we suggest our experimental colleagues to
explore the heavy baryon-antibaryon production at B- and charm-tau factories.

From eqs. (\ref{baryon}) we can see that as s is sufficiently large the cross
section beside the factor $|\xi (-\omega)|^2$ will increase with s. However,
unitarity requires that the cross section decrease in the end. We believe
that this can be achieved by the behaviour of $\xi (-\omega)$ at large
$\omega$. As mentioned above $\xi (-\omega)$ is not evaluated up to now.
In the time-like region several models \cite{guok}, \cite{jenkins} suggest
that $\xi (\omega)$ is suppressed by exponential or higher order $1/\omega$ 
factors as $\omega$ increases.

Our conclusion is that in terms of the superflavor symmetry, we evaluate the
ratio of the 
production rates of baryon-antibaryon pair which contains two heavy quarks
(antiquarks) and that of the meson-antimeson
(for example $D\overline D$). It is found that this ratio is $10^{-5}$
to $10^{-8}$. One can be optimistic to the measurements in the
proposed B- and charm-tau factories.

\vspace{1.5cm}

\noindent {\bf Acknowledgment}

Two of us (Jin and Li) would like to thank the International Center for Theoretical
Physics which provides them with a wonderful working atmosphere and
excellent library, and part of the work is done over there.
One of us (Guo) is indebted to the Institute of Physics for the hospitality
during his stay. Li is also
indebted to Prof. C. Chang, Dr. Y. Chen and G. Han for helpful discussions.

\vspace{4cm}

\pagebreak
\noindent{Appendix A}

\vspace{0.5cm}

\noindent { The differential cross section $d\sigma/dcos\theta$ 
for $e^+e^-\rightarrow X_s\bar X_{s'}$}
\vspace{1cm}

\noindent (i) For $e^+e^-\rightarrow X_S\overline X_S$,

\begin{equation}
{d\sigma\over dcos\theta} = {1\over 8\pi s^3\sqrt s}
e^2g_S^2 |\xi(-\omega)|^2 {1\over 8M^2}s(s-4M^2)^{5/2}(1-cos^2\theta).
\end{equation}

\vspace{0.6cm}

\noindent (ii) For $e^+e^-\rightarrow X_A^{1/2}\overline X_A^{1/2}$

\begin{eqnarray}
{d\sigma\over dcos\theta} &=& {e^2g_A^2\over
4608\pi s^2\sqrt s M^6}\sqrt{(s-4M^2)}|\xi(-\omega)|^2
[4(16M^4-8M^2 s+s^2)(1+cos^2 \theta) + \nonumber \\
&& (64(2a+1)^2M^8+32(-8a^2-2a+1)M^6 s+12(8a^2-4a-1)M^4s^2+4(-4a^2+5a-1)M^2s^3
\nonumber \\
&& +(a-1)^2 s^4)(1-cos^2 \theta)].
\end{eqnarray}

\vspace{0.6cm}

\noindent (iii) For $e^+e^-\rightarrow X_A^{3/2}\overline X_A^{3/2}$,

\begin{eqnarray}
{d\sigma\over dcos\theta} &=& {e^2g_A^2\over
2304\pi s^2\sqrt s M^4}\sqrt{(s-4M^2)}|\xi(-\omega)|^2
[s(64M^4-28M^2 s+3s^2)(1+cos^2 \theta) + \nonumber \\
&& 2(288a^2M^8+16(-11a^2+3a)M^6 s+2(25a^2-26a+8)M^4 s^2+2(-5a^2+9a-4)M^2 s
\nonumber \\
&& +(a-1)^2 s^4)(1-cos^2 \theta)].
\end{eqnarray}

\vspace{0.6cm}

\noindent (iv) For $e^+e^-\rightarrow X_A^{1/2}\overline X_A^{3/2}$,

\begin{eqnarray}
{d\sigma\over dcos\theta} &=& {e^2g_A^2\over
2304\pi s\sqrt s M^4}\sqrt{(s-4M^2)}|\xi(-\omega)|^2
[(16M^4-8M^2 s+s^2)(1+cos^2 \theta) + \nonumber \\
&& (16(a+1)^2M^4-8(a-1)^2M^2 s+(a-1)^2 s^2)(1-cos^2 \theta)].
\end{eqnarray}

\vspace{4cm}
\pagebreak

\noindent{Appendix B}

\centerline {Table 2 Cross section for scalar diquark
($10^{-16}$ GeV$^{-2}$$|\xi(-\omega)|^2)$}

\begin{center}

\begin{tabular}{|r|r|r|r|r|r|r|r|r|}
\hline
$\omega$ &1.0 &1.1 &1.2& 1.3 &1.4 &1.6&1.8&2.0 \\
\hline
$\sqrt{s}(GeV)$ & 13.8 & 14.1 & 14.5 & 14.8 &15.1  &15.7  &16.3&16.9\\
\hline
$\sigma (e^+e^- \rightarrow X_S\overline X_S)$ & 0 & 2.0 & 7.6 & 14.0& 19.0 &23.2
&20.1 &15.4\\
\hline
\end{tabular}

\end{center}

\vspace{0.2cm}
\centerline {Table 3 Cross section for baryon containing axial cc diquark
($10^{-13}$ GeV$^{-2}$$|\xi(-\omega)|^2)$}

\begin{center}

\begin{tabular}{|r|r|r|r|r|r|r|r|r|}
\hline
$\omega$ &1.0 &1.1 &1.2& 1.3 &1.4 &1.6&1.8&2.0 \\
\hline
$\sqrt{s}(GeV)$ &7.14 & 7.32 & 7.49 & 7.66 &7.82  &8.14  &8.45&8.75\\
\hline
$\sigma (e^+e^- \rightarrow X_{A,1/2} \overline X_{A, 1/2})$ & 0 & 0.6 & 2.8 
& 6.0& 9.7 &17.0 &23.1 &27.5\\
\hline
$\sigma (e^+e^- \rightarrow X_{A,1/2} \overline X_{A, 3/2})$ & 0 & 0.2 & 1.0 
& 2.3& 3.8 &6.9 &9.7 &11.9\\
\hline
$\sigma (e^+e^- \rightarrow X_{A,3/2} \overline X_{A, 3/2})$ & 0 & 1.3 & 5.8 
& 12.5& 20.3 &36.1 &49.5 &59.4\\
\hline
\end{tabular}

\end{center}

\vspace{0.2cm}

\centerline {Table 4 Cross section for baryon containg axial bc diquark
($10^{-14}$ GeV$^{-2}$$|\xi(-\omega)|^2)$}

\begin{center}

\begin{tabular}{|r|r|r|r|r|r|r|r|r|}
\hline
$\omega$ &1.0 &1.1 &1.2& 1.3 &1.4 &1.6&1.8&2.0 \\
\hline
$\sqrt{s}(GeV)$ & 13.8 & 14.1 & 14.5 & 14.8 &15.1  &15.7  &16.3&16.9\\
\hline
$\sigma (e^+e^- \rightarrow X_{A,1/2} \overline X_{A, 1/2})$ & 0 & 0.2 & 0.8 
& 1.7& 2.7 &4.6 &6.1 &7.1\\
\hline
$\sigma (e^+e^- \rightarrow X_{A,1/2} \overline X_{A, 3/2})$ & 0 & 0.05 & 0.2 
& 0.4& 0.7 &1.2 &1.7 &2.0\\
\hline
$\sigma (e^+e^- \rightarrow X_{A,3/2} \overline X_{A, 3/2})$ & 0 & 0.4 & 1.8 
& 3.8& 6.1 &10.5 &13.9 &16.2\\
\hline
\end{tabular}

\end{center}  

\vspace{0.2cm}

\centerline {Table 5 Cross section for baryon containing axial bb diquark
($10^{-16}$ GeV$^{-2}$$|\xi(-\omega)|^2)$}

\begin{center}

\begin{tabular}{|r|r|r|r|r|r|r|r|r|}
\hline
$\omega$ &1.0 &1.1 &1.2& 1.3 &1.4 &1.6&1.8&2.0 \\
\hline
$\sqrt{s}(GeV)$ & 20.3 & 20.8  &21.3 & 21.8 & 22.3 &23.2  &24.1&24.9\\
\hline
$\sigma (e^+e^- \rightarrow X_A,1/2 \overline X_A, 1/2)$ & 0 & 0.3 & 1.4 
& 3.0& 4.8 &8.5 &11.5 &13.6\\
\hline
$\sigma (e^+e^- \rightarrow X_A,1/2 \overline X_A, 3/2)$ & 0 & 0.1& 0.5 
& 1.1 & 1.9 & 3.4 & 4.8 & 5.9\\
\hline
$\sigma (e^+e^- \rightarrow X_A,3/2 \overline X_A, 3/2)$ & 0 & 0.6 & 2.9 
& 6.2& 10.1 &17.9 &24.6 &29.5\\
\hline
\end{tabular}

\end{center}  

\vspace{0.2cm}

\centerline {Table 6 Cross section for $D \overline D$
($10^{-8}$ GeV$^{-2}$$|\xi(-\omega)|^2)$}

\begin{center}

\begin{tabular}{|r|r|r|r|r|r|r|r|r|}
\hline
$\omega$ &1.0 &1.1 &1.2& 1.3 &1.4 &1.6&1.8&2.0 \\
\hline
$\sqrt(s)(GeV)$ & 3.7 & 3.8 & 3.9 & 4.0  &4.1  &4.3 &4.4 &4.6\\
\hline
$\sigma (e^+e^- \rightarrow D\overline D)$ & 0 & 1.8 & 4.4 & 7.3 & 10.1 &15.1
&19.3 &22.8\\
\hline
\end{tabular}

\end{center}

\pagebreak

\noindent {\bf Figure Captions}

Fig.1, The leading Feynman diagrams where the emitted gluon as the intermediate
boson is hard and the form factor of the effective vertices are calculated
in the framework of perturbative QCD.\\

\end{document}